\documentclass[superscriptaddress,twocolumn,showpacs,preprintnumbers,amsmath,amssymb, prx, reprint, longbibliography]{revtex4-1}
\usepackage{amsmath}
\usepackage{stmaryrd}
\usepackage{txfonts}
\usepackage{amssymb}
\usepackage{mathrsfs}
\usepackage{graphicx}
\usepackage{dcolumn}
\usepackage{bm}
\usepackage{epsfig}
\usepackage{color}
\usepackage{pbox}
\usepackage[colorlinks=true, linkcolor=blue, citecolor=blue, urlcolor=blue]{hyperref} 
\begin{document}

\title{Intrinsic left-handed electromagnetic properties in anisotropic superconductors}

\author{Shi-Zeng Lin}
\email{szl@lanl.gov}
\affiliation{Theoretical Division, T-4 and CNLS, Los Alamos National Laboratory, Los Alamos, New Mexico 87545, USA}
\author{Hou-Tong Chen}
\affiliation{Center for Integrated Nanotechnologies, Los Alamos National Laboratory, Los Alamos,
New Mexico 87545, USA}

\begin{abstract}
Left-handed materials usually are realized in artificial subwavelength structures. Here we show that some anisotropic superconductors, such as $\mathrm{Bi_2Sr_2CaCu_2O_{8+\delta}}$, $\mathrm{YBa_2Cu_xO_y}$ and $\mathrm{La_{2-x}Sr_xCuO_4}$, are intrinsic left-handed materials. The condition is that the plasma frequency in the $c$ axis, $\omega_c$, and in the $ab$ plane, $\omega_{ab}$, and the operating frequency, $\omega$, satisfy $\omega_c<\omega<\omega_{ab}$. In addition $\omega$ should be smaller than the superconducting energy gap to sustain superconductivity. We study the reflection and transmission of electromagnetic waves, and reveal negative refraction and backward wave with phase velocity opposite to the direction of energy flux propagation. We also discuss possible approaches of improvement, making these properties feasible for experimental validation. Being intrinsic left-hand materials, the anisotropic superconductors are promising for applications in novel electromagnetic devices in the terahertz frequency band.

\end{abstract}
\date{\today}
\maketitle

\begin{figure}[t]
\psfig{figure=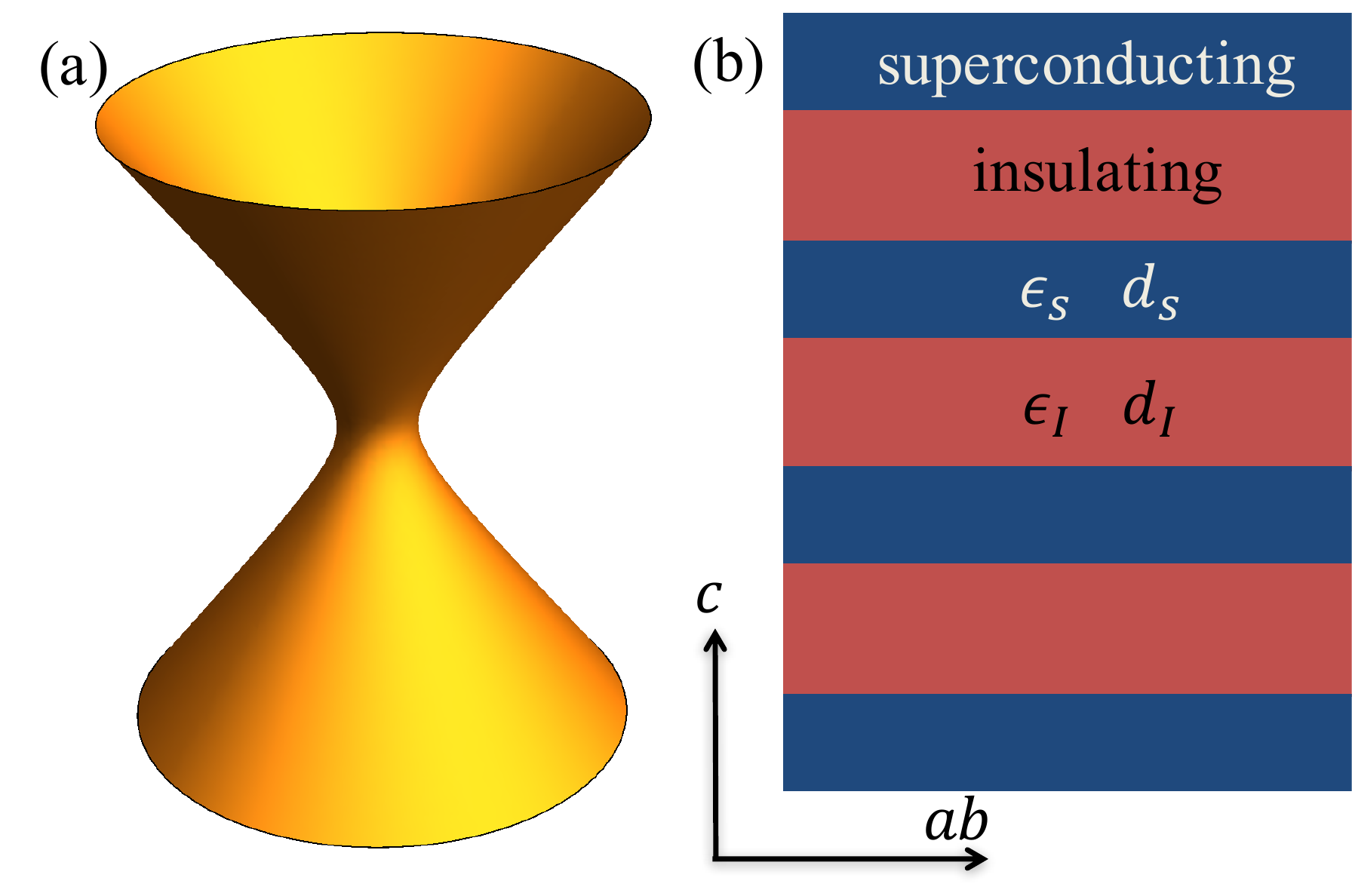,width=\columnwidth}
\caption{(color online) (a) Isofrequency surfaces given by $\omega_3(\mathbf{q})=\mathrm{constant}$ for the plasma mode in Eq. \eqref{eq9}. (b) Schematic view of a superconductor/insulator superlattice.} \label{f1}
\end{figure}

In conventional materials or the right-handed materials, the group velocity and phase velocity of electromagnetic waves are in the same direction. In contrast, the group velocity is opposite to the phase velocity in left-handed medium (LHM). \cite{Veselago1968} Because of their novel electromagnetic properties, LHM are promising for many applications, including antennas, superlens and cloaking. \cite{shalaev_optical_2007,chen_review_2016} To realize the LHM, generally one needs to artificially engineer the electromagnetic medium in subwavelength scale to create superlattice structures with effective permittivity and permeability. \cite{PhysRevLett.84.4184,shelby_experimental_2001} These electromagnetic superlattices are metamaterials which are currently under active research. There is a family of metamaterials with a hyperbolic dispersion of the electromagnetic waves, called hyperbolic metamaterials. \cite{PhysRevLett.90.077405,belov_backward_2003,poddubny_hyperbolic_2013,cortes_quantum_2012} The isofrequency surface of the wave vector is hyperbolic, see Fig. \ref{f1}(a), instead of an ellipsoid in conventional materials. One way to realize the hyperbolic electromagnetic dispersion is to use layered metal-dielectric periodic structures. \cite{agranovich_notes_1985} The dielectric constant in metals at frequency below the plasma frequency is negative while the dielectric constant for the dielectric medium is positive. By properly choosing the thickness of the metallic and dielectric layers, the layered superlattice can have a negative effective dielectric constant for electromagnetic waves propagating in one direction. To date, most LHMs are artificial structure, while the naturally occurring LHMs are limited. \cite{yoxall_direct_2015}

Anisotropic layered crystals are ubiquitous. One notable class of layered materials are superconductors, such as cuprate and iron pnictide superconductors. Here we show that the anisotropic superconductors are possible intrinsic LHMs, and find the conditions for realizing the LHMs. We have already identified $\mathrm{Bi_2Sr_2CaCu_2O_{8+\delta}}$
(BSCCO), $\mathrm{YBa_2Cu_xO_y}$ (YBCO)  and $\mathrm{La_{2-x}Sr_xCuO_4}$ (LSCO) cuprate superconductors as LHMs. In analog to metals, superconductors also host plasma mode which is a collective excitation of Cooper pairs and electromagnetic fields. Because of the weak interlayer superconducting coupling in anisotropic superconductors \cite{Kleiner92,PhysRevB.49.1327}, the plasma frequency is anisotropic and the plasma frequency in the $c$ axis can be as low as sub-terahertz and blew the superconducting energy gap. \cite{Tachiki94,Bulaevskii94} Because the quasiparticle excitations are gapped at the operating frequencies, the dissipation caused by quasiparticles can be minimized. The upper bound of the operating frequency is limited by the superconducting energy gap. For high $T_c$ cuprate superconductors, the upper frequency limit can be as high as 10 THz. Therefore the superconducting LHMs can operate in the terahertz frequency band, a frequency band is noticeably underutilized but has huge potential for applications. \cite{Ferguson02,Tonouchi07} For highly anisotropic superconductors where the superconducting layers are coupled through weak Josephson interaction, the inductance of the system can be tuned by bias current or external magnetic field, therefore allowing for tuning of the electromagnetic (EM) response. All these unique properties make anisotropy superconductors promising for applications in EM devices with left-hand properties.  


We consider an anisotropic superconductor with in-plane ($ab$ plane) plasma frequency, $\omega_{ab}=c/\lambda_{ab}$, above the superconducting energy gap,  $\Delta$, and the out-of-plane ($c$ axis) plasma frequency, $\omega_{c}=c/\lambda_c$, below $\Delta$, i.e. $\omega_{c}<\Delta<\omega_{ab}$. Here $\lambda_{ab}$ and $\lambda_{c}$ are the London penetration depth. Examples are BSCCO \cite{Kleiner92,PhysRevB.49.1327}, YBCO \cite{PhysRevB.50.3511,PhysRevLett.89.247001}, LSCO \cite{PhysRevLett.69.1455,PhysRevLett.72.2263} and some organic superconductors \cite{PhysRevB.55.R11977,PhysRevB.62.5965}. The length scales that are relevant for EM properties are the London penetration depth, which are much bigger than the superconducting coherence length and the crystal lattice constant. We adopt the anisotropic Ginzburg-Landau equation and assuming the amplitude of the superconducting order parameter $\Psi=|\Psi|\exp(i\phi)$ does not change in space and time, valid for weak incident waves with frequency below the superconducting energy gap. The Ginzburg-Landau free energy functional for superconductivity is  
\begin{equation}\label{eq1}
\mathcal{F}= {{\alpha_T}|{\Psi }{|^2} + \frac{{{\beta}}}{2}|{\Psi }{|^4} +\sum_{\mu=x, y, z}\frac{\hbar^2}{{2{m_\mu}}}{{\left| {\left(-i\partial_\mu-\frac{2\pi A_\mu}{\Phi_0}\right){\Psi}} \right|}^2}},
\end{equation}
where the anisotropy is due to the anisotropic Cooper pairs mass $m_\mu$. Here $\Phi_0=hc/2e$ is the quantum flux. The supercurrents in the $ab$ plane $J_{ab}$ and the $c$ axis $J_{c}$ are obtained by taking derivative of $\mathcal{F}$ with respect to $A_{ab}$ and $A_c$
\begin{equation}\label{eq2}
J_{{\mu}}=\frac{c \Phi _0 }{8 \pi ^2 \lambda _{\mu}^2}\left(\partial_\mu \phi -\frac{2 \pi  A_\mu}{\Phi _0}\right),
\end{equation}
where $\mu=x,\ y,\ z$ and $\lambda_x=\lambda_y=\lambda_{ab}$, $\lambda_z=\lambda_c$. The time variation of the supercurrent generates electric fields, $\mathbf{E}$, according to the London equation
\begin{equation}\label{eq3}
\partial_t J_{\mu}=\frac{c^2}{4 \pi \lambda _{\mu}^2} E_\mu.
\end{equation}
For the Josephson-coupled superconductors, Eq. \eqref{eq3} for $J_c$ needs to be modified to account for the gauge invariant phase $\phi$. \cite{Bulaevskii94,Koshelev01} Here we take $\phi=0$ valid when no external current or magnetic field is applied. We will discuss the effect of $\phi$ below. Using the Ampere's law, we have
\begin{equation}\label{eq4}
\left(\nabla\times\mathbf{B}\right)_\mu=\frac{4\pi}{c}\left(J_{\mu}+\sigma_\mu E_\mu\right)+\frac{1}{c}\partial_t E_\mu\equiv \frac{\epsilon_\mu}{c}\partial_t E_\mu,
\end{equation}
where at the right-hand side are the total current density consisting of supercurrent, normal dissipative current with conductivity $\sigma_\mu$ and the displacement current. The dielectric functions in the frequency domain $\epsilon_\mu(\omega)$ is 
\begin{equation}\label{eq5}
\epsilon_\mu(\omega)=1-\frac{\omega_{\mu}^2}{\omega^2}-\frac{i4\pi\sigma_{\mu}}{\omega}.
\end{equation}
Here $\epsilon_\mu(\omega)$ has imaginary part due to the dissipation. In the case $\omega_c<\Delta<\omega_{ab}$ considered here, $\mathrm{Re}[\epsilon_{c}]>0$ and $\mathrm{Re}[\epsilon_{ab}]<0$ in the frequency region $\omega_c <\omega < \Delta$, which are essential for the following results. 

First let us study the plasma modes inside an anisotropic superconductor. Using the Faraday's law $\nabla\times \mathbf{E}=-\partial_t\mathbf{B}/c$ and Eq. \eqref{eq4}, we obtain equation for $\mathbf{E}$ in the Fourier space $\mathbf{E}(\mathbf{r}, t)\sim \mathbf{E}(\mathbf{q}, \omega)\exp[i (\mathbf{q}\cdot\mathbf{r}+\omega t)]$
\begin{equation}\label{eq6}
\left( {\begin{array}{*{20}{c}}
{q_y^2 + q_z^2}&{ - {q_x}{q_y}}&{ - {q_z}{q_x}}\\
{ - {q_x}{q_y}}&{q_z^2 + q_x^2}&{ - {q_y}{q_z}}\\
{ - {q_z}{q_x}}&{ - {q_y}{q_z}}&{q_x^2 + q_y^2}
\end{array}} \right)\left( {\begin{array}{*{20}{c}}
{{E^x}}\\
{{E^y}}\\
{{E^z}}
\end{array}} \right) = \frac{{{\omega ^2}}}{{{c^2}}}\left( {\begin{array}{*{20}{c}}
{{\epsilon_{ab} E^x}}\\
{{\epsilon_{ab} E^y}}\\
{{\epsilon_{c} E^z}}
\end{array}} \right).
\end{equation}
In our notation, the phase velocity is $\mathbf{v}\cdot \mathbf{q}=-\omega$. There are three plasma modes with the eigen frequencies and eigen vectors
\begin{equation}\label{eq7}
\frac{\omega_1^2}{c^2}=0, \ \mathbf{E}_1=[q_x,\ q_y,\ q_z],
\end{equation}
\begin{equation}\label{eq8}
\frac{\omega_2^2}{c^2}=\frac{q_x^2+q_y^2+q_z^2}{\epsilon_{ab}}, \ \mathbf{E}_2=[-q_y,\ q_x,\ 0],
\end{equation}
\begin{equation}\label{eq9}
\frac{\omega_3^2}{c^2}=\frac{q_x^2+q_y^2}{\epsilon_{c}}+\frac{q_z^2}{\epsilon_{ab}}, \ \mathbf{E}_3=[\epsilon_{c} q_x q_z,\ \epsilon_{c} q_y q_z,\ -\epsilon_{ab} (q_x^2+q_y^2)].
\end{equation}
The first mode is a static mode. The second mode is the usual EM mode in the right-handed materials. Because $\mathrm{Re}[\epsilon_{c}]>0$ and $\mathrm{Re}[\epsilon_{ab}]<0$, the isofrequency surface of $\omega_3$ is hyperbolic as shown in Fig. \ref{f1}(a). The third mode $\omega_3$ is the nontrivial plasma mode giving rise to the left-handed properties, and we will focus on $\omega_3$ in the following discussions. We will demonstrate below that anisotropic superconductors are intrinsic LHMs by considering the reflection and refraction of EM waves \cite{belov_backward_2003} at its $bc$ and $ab$ surfaces.

\begin{figure}[b]
\psfig{figure=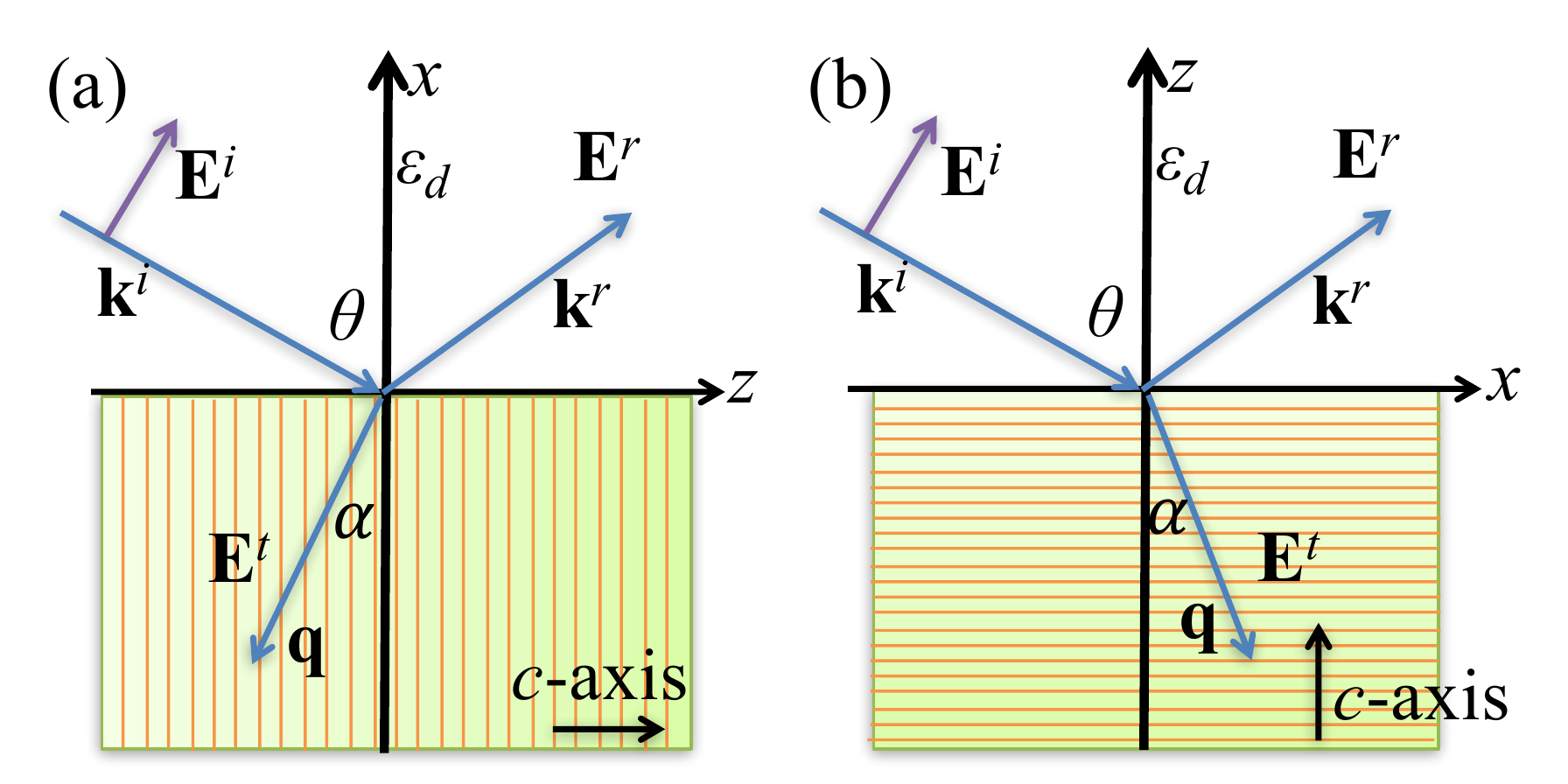,width=\columnwidth}
\caption{(color online) Reflection and transmission of EM wave at the (a) $bc$ surface (b) $ab$ surface of an anisotropic superconductor. The blue lines with arrows represent the energy flux flow.
} \label{f2}
\end{figure}


First we study reflection of transverse magnetic wave at the $bc$ surface as displayed in Fig. \ref{f2}(a), where only the $\omega_3$ mode is excited. The system is uniform in the $y$ axis and $k_y=q_y=0$. In the upper region with an dielectric constant $\epsilon_d$, the dispersion of the EM wave is given by $(k_x^2+k_z^2)/{\epsilon_d}=\omega^2/c^2$.  At the interface $k_z$ and $q_z$ are continuous, and we have $q_z = k_z = - \sqrt{\epsilon_d}\omega\sin\theta/c$, where the negative sign originates from the phasor definition of EM waves and indicates wave propagation in the positive $z$ direction. From Eq. \eqref{eq9}, we obtain $q_x=\pm \sqrt{\epsilon_{c}(1-\epsilon_d\sin^2\theta/\epsilon_{ab})}\omega/c$. We will determine the sign of $q_x$ from the direction of the energy flux which has a component in the negative $z$ direction. The imaginary part of $q_x$ accounts for the dissipation. We consider the case where dissipation is weak as in the case of BSCCO, $\mathrm{Im}[\epsilon_\mu]\ll \mathrm{Re}[\epsilon_\mu]$, and neglect the dissipation in the following derivations. Using the boundary conditions at the interface, $(E^i-E^r)\cos\theta=E_z^t$ and $\epsilon_d(E^i+E^r)\sin\theta=\epsilon_{ab}E_x^t$, we obtain the electric field for the reflected and transmitted waves, 
\begin{equation}\label{eq10}
E^r=\frac{q_z\epsilon_{c}\cos\theta +q_x\epsilon_d\sin\theta}{q_z\epsilon_{c}\cos\theta -q_x\epsilon_d\sin\theta}E^i,
\end{equation}
\begin{equation}\label{eq11}
E_x^t=-\frac{\epsilon_c}{\epsilon_{ab}}\frac{q_z\epsilon_{d}\sin(2\theta)}{q_x\epsilon_d\sin\theta-q_z\epsilon_{c}\cos\theta }E^i,
\end{equation}
\begin{equation}\label{eq12}
E_z^t=\frac{q_x\epsilon_{d}\sin(2\theta)}{q_x\epsilon_d\sin\theta-q_z\epsilon_{c}\cos\theta }E^i.
\end{equation}
We then consider the energy flow inside the superconductor by calculating the Poynting vector $\mathbf{P}=\frac{c}{8\pi}\mathrm{Re}[\mathbf{E}\times \mathbf{B}^*]=-\frac{c^2}{8\pi\omega}\mathrm{Re}[\mathbf{E}\times (\mathbf{q}^*\times\mathbf{E}^*)]$,
\begin{equation}\label{eq13}
\mathbf{P}=-\frac{\omega \epsilon_c}{8\pi}\left(q_x\hat{x}+\frac{\epsilon_c}{\epsilon_{ab}}q_z\hat{z}\right)\left(\frac{\epsilon_d\sin{(2\theta)}|E^i|}{q_x\epsilon_d\sin\theta-q_z\epsilon_{c}\cos\theta }\right)^2,
\end{equation}
where $\hat{x}$ and $\hat{z}$ are unit vectors in the $x$ and $z$ directions, respectively. Here $\mathbf{P}\cdot\hat{x}<0$ must be negative because the energy flow into the negative $x$ direction, which requires $q_x>0$. Therefore the $x$ component of the phase velocity and the energy flux flow are in the same direction. Because $q_z<0$ and $\epsilon_{ab}<0$, the $z$ component of the energy flux flow, $\mathbf{P}\cdot\hat{z}<0$, suggests the opposite direction to the phase velocity. This \emph{negative refraction} has the refraction angle
\begin{equation}\label{eq14}
\tan\alpha\equiv \frac{\mathbf{P}\cdot\hat{z}}{\mathbf{P}\cdot\hat{x}}=\frac{\sqrt{\epsilon_d\epsilon_{c}}\sin\theta}{\sqrt{\epsilon_{ab}^2-\epsilon_d\epsilon_{ab}\sin^2\theta}}.
\end{equation}
The reflection coefficient is
\begin{equation}\label{eq15}
R=\frac{E^r}{E^i}=\frac{\cos\theta\sqrt{\epsilon_{c}}-g\sqrt{\epsilon_d}}{\cos\theta\sqrt{\epsilon_{c}}+g\sqrt{\epsilon_d}},\ \ g\equiv \sqrt{1-\frac{\epsilon_d}{\epsilon_{ab}}\sin^2\theta}.
\end{equation}
There is always reflection because $\epsilon_c<1\le\epsilon_d$ according to Eq. \eqref{eq5}. The transmission coefficient is
\begin{equation}\label{eq16}
T=\frac{E^t}{E^i}=\sqrt{\frac{\epsilon_c\epsilon_d}{\epsilon_{ab}^2}\sin^2\theta+g^2}\frac{2\sqrt{\epsilon_d}\cos\theta}{\cos\theta\sqrt{\epsilon_{c}}+g\sqrt{\epsilon_d}}.
\end{equation}
At the BSCCO and vacuum interface, $\mathrm{Re}[\epsilon_c]<1$, $\epsilon_d=1$ and $-\mathrm{Re}[\epsilon_{ab}]\gg 1$, substantial portion of the incident wave can still transmitted into the $bc$ surface of the BSCCO with a small refraction angle $\alpha\propto 1/|\epsilon_{ab}|\ll 1$. Basically the transmitted wave travels along the layer direction with the electric polarization in the crystal $c$ axis because of the strong anisotropy in BSCCO \cite{Tachiki94,Bulaevskii94}.

The EM waves transmitted into BSCCO decay because of the dissipation. For $\omega_c\approx 1.5\ \mathrm{THz}$, $\omega_{ab}=750\ \mathrm{THz}$, $\sigma_{ab}\approx 4\times 10^6\ \mathrm{\Omega\cdot m}^{-1}$ and $\sigma_{c}\approx 0.2\ \mathrm{\Omega\cdot m}^{-1}$ in BSCCO \cite{Latyshev1999,Corson2000}, the decay length is about $r _d\approx 2\ \mathrm{cm}$ at an incident angle $\theta=\pi/4$ and operating frequency $\omega=5\omega_c$ and $\epsilon_d = 1$, which would be simple to perform transmission experiments using bulk BSCCO crystals. However, it would be technically challenging to experimentally measure the small negative refraction angle $\alpha$ in highly anisotropic superconductors using the configuration in Fig. \ref{f2}(a). For BSCCO, we estimate $\alpha\approx \sin\theta/10000$ at operating frequency $\omega=5\omega_c$. Therefore, one only measures the wave propagation along the $ab$ plane, resulting a trivial beam shift when a BSCCO slab is used. To achieve a sizable negative refraction angle, one may use less anisotropic materials with $\omega_c<\omega<\Delta<\omega_{ab}$. One may introduce defects into superconductors to increase London penetration depth and to render the superconductor less anisotropic, or use superconductor/dielectric multilayer artificial structures where the anisotropic properties can be tuned through tailoring the layer thicknesses, see the discussions below. 

Let us proceed to study the reflection and transmission of EM wave at the $ab$ plane of an anisotropic superconductor, as displayed in Fig. \ref{f2}(b). In this case, we have $q_z=-\sqrt{\epsilon_{ab}(1-\epsilon_d\sin^2\theta/\epsilon_c)}\omega/c$. We have taken negative sign here because the energy flux flow in Eq. \eqref{eq20} must be in the negative $z$ direction. The incident angle $\theta$ should satisfy $\sin\theta\ge \sqrt{\epsilon_c/\epsilon_d}$ to have refraction. Using the boundary conditions, $(E^i-E^r)\cos\theta=E_x^t$ and $\epsilon_d(E^i+E^r)\sin\theta=\epsilon_{c}E_z^t$, we obtain
\begin{equation}\label{eq17}
E^r=\frac{q_x\epsilon_{ab}\cos\theta +q_z\epsilon_d\sin\theta}{q_x\epsilon_{ab}\cos\theta -q_z\epsilon_d\sin\theta}E^i,
\end{equation}
\begin{equation}\label{eq18}
E_x^t=\frac{q_z\epsilon_{d}\sin(2\theta)}{q_z\epsilon_d\sin\theta-q_x\epsilon_{ab}\cos\theta }E^i,
\end{equation}
\begin{equation}\label{eq19}
E_x^t=-\frac{\epsilon_{ab}}{\epsilon_{c}}\frac{q_x\epsilon_{d}\sin(2\theta)}{q_z\epsilon_d\sin\theta-q_x\epsilon_{ab}\cos\theta }E^i.
\end{equation}
The Poynting vector is
\begin{equation}\label{eq20}
\mathbf{P}=-\frac{\omega \epsilon_{ab}}{8\pi}\left(\frac{\epsilon_{ab}}{\epsilon_{c}}q_x\hat{x}+q_z\hat{z}\right)\left(\frac{\epsilon_d\sin{(2\theta)}|E^i|}{q_z\epsilon_d\sin\theta-q_x\epsilon_{ab}\cos\theta }\right)^2.
\end{equation}
The required negative $q_z$ (i.e., phase velocity in positive $z$ direction) suggests that the $z$ component of the EM energy flux flow is opposite to that of the phase velocity, i.e., \emph{backward waves}. We also have $\hat{x}\cdot\mathbf{P}>0$ because $q_x=-\sqrt{\epsilon_d}\omega\sin\theta/c<0$, therefore we have normal refraction at the interface. The refraction angle is
\begin{equation}\label{eq21}
\tan\alpha\equiv -\frac{\mathbf{P}\cdot\hat{x}}{\mathbf{P}\cdot\hat{z}}=\frac{\sqrt{-\epsilon_d\epsilon_{ab}}\sin\theta}{\sqrt{\epsilon_d\epsilon_{c}\sin^2\theta-\epsilon_{c}^2}}.
\end{equation}
The reflection and transmission coefficients are
\begin{equation}\label{eq22}
R=\frac{\cos\theta\sqrt{-\epsilon_{ab}}-g_c\sqrt{\epsilon_d}}{\cos\theta\sqrt{-\epsilon_{ab}}+g_c\sqrt{\epsilon_d}},\ \ g_c\equiv \sqrt{\frac{\epsilon_d}{\epsilon_{c}}\sin^2\theta-1},
\end{equation}
\begin{equation}\label{eq23}
T=\frac{E^t}{E^i}=\sqrt{g_c^2-\frac{\epsilon_{ab}\epsilon_d}{\epsilon_{c}^2}\sin^2\theta}\frac{2\sqrt{\epsilon_d}\cos\theta}{\cos\theta\sqrt{-\epsilon_{ab}}+g_c\sqrt{\epsilon_d}}.
\end{equation}

For BSCCO or LSCCO, only a small fraction of the incident EM can penetrate into the superconductor from the $ab$ surface because the refraction angle is $\alpha\approx \pi/2$ for a not small incident angle $\theta$ and the transmitted wave travels along the interface, or the incident wave is almost completely reflected, $R\approx 1$, for a small $\theta$. It is difficult for EM to propagate in the crystal $c$ axis in the frequency region $\omega<\omega_{ab}$ because the EM wave has to cross the superconducting layers. The EM wave can travel along the weak superconducting region in the $ab$ plane with plasma frequency $\omega_c$. In order to realize significant transmission, one needs to reduce the the effective $\epsilon_{ab}$, either using less anisotropic superconductors or using superconductor/dielectric artificial structures, where one may be able to measure the opposite directions of phase and group velocity using terahertz time-domain spectroscopy \cite{Grischkowsky90}.

One of the advantages using anisotropy superconductors such as BSCCO is that the superconducting layers are coupled through the Josephson junctions, where the plasma frequency $\omega_c$ can be tuned by the bias current in the $c$ axis or external magnetic field in the $ab$ plane. In this case, we need to replace Eq. \eqref{eq3} by $\partial_t J_c=c^2 E_z\langle\cos\phi\rangle/(4\pi\lambda_c^2)$, where $\langle\cdots\rangle$ represents spatial average. In the presence of a bias current $I_{\mathrm{ext}}$, $\phi$ is given by $\langle\sin\phi\rangle=I_{\mathrm{ext}}/I_{c0}$ with $I_{c0}$ the critical current. We have $\omega_c=c^2\cos\phi/\lambda_c^2$. This points a way to tune $\epsilon_c$ and to achieve controlled functionalities of LHMs by applying bias current or magnetic field. 

As previously mentioned, in superconductors such as BSCCO the highly anisotropy results in minimal negative refraction angle or transmission. Here we discuss an alternative realization of superconducting LHMs with controlled EM response by using artificial superconductor/dielectric superlattice as depicted in Fig. \ref{f1}(b). Such a superlattice can be modeled as a stack of Josephson junctions, which have been discussed extensively in the past both experimentally \cite{PhysRevB.66.064527,goldobin_cherenkov_2000} and theoretically \cite{sakai_fluxons_1993,PhysRevB.50.6919}. We employ the effective medium theory to obtain the dielectric constant of the superlattice. We consider that the electromagnetic wavelength is much bigger than the period of the superlattice and neglect the variation of EM fields inside the layers. The electric displacement field $\mathbf{D}_{l}$ in the $l$-th superconducting layer is $\mathbf{D}_{s, l}=\epsilon_{s}\mathbf{E}_{s, l}$ and similarly for the insulating layer $\mathbf{D}_{I, l}=\epsilon_{I}\mathbf{E}_{s, l}$. Here we consider an isotropic superconductor with $\epsilon_s=1-\omega_s^2/\omega^2-i4\pi\sigma/\omega$, where the plasma frequency is $\omega_s=c/\lambda$ with $\lambda$ the London penetration depth and $\sigma$ the electric conductivity. Averaging $\mathbf{D}$ and $\mathbf{E}$ over one period and using the standard EM boundary condition, we obtain $D_{ab}\equiv (d_sD_{s,ab}+d_I D_{I,ab})/(d_s+d_I)=(d_s\epsilon_s+d_I\epsilon_I)/(d_s+d_I)E_{l,ab}$ and $E_{c}\equiv (d_s E_{s,c}+d_I E_{I,c})/(d_s+d_I)=(d_s\epsilon_s^{-1}+d_I\epsilon_I^{-1})D_c/(d_s+d_I)$. The effective permittivity are \cite{agranovich_notes_1985}
\begin{equation}\label{eq24}
\epsilon_{ab}=\frac{d_s\epsilon_s+d_I\epsilon_I}{d_s+d_I}, \ \ \epsilon_{c}=(d_s+d_I)\left(\frac{d_s}{\epsilon_s}+\frac{d_I}{\epsilon_I}\right)^{-1}.
\end{equation}
For $\omega<\omega_s$, $\mathrm{Re}[\epsilon_s]<0$. The anisotropy ratio $\epsilon_c/\epsilon_{ab}$ can be tuned by the thickness $d_s$ and $d_I$. Moreover $\epsilon_c$ can be tuned by bias current and magnetic field in the presence of Josephson coupling, when the superconducting coherence length is larger than the thickness of the dielectric layers, where it would be possible to achieve active tuning by applying electrical currents or magnetic field, similar to the case of BSCCO.


To summarize, we demonstrate that for anisotropic superconductors with plasma frequency in the $c$ axis, $\omega_c$, and in the $ab$ plane, $\omega_{ab}$, they are intrinsic left-handed materials in the operating frequency region $\omega_c<\omega<\omega_{ab}$. Meanwhile $\omega$ should be below the superconducting energy gap in order to maintain superconductivity. We study the reflection and transmission of electromagnetic waves at the surface of superconductors and found the existence of negative refraction and backward waves. We identify $\mathrm{Bi_2Sr_2CaCu_2O_{8+\delta}}$, $\mathrm{YBa_2Cu_xO_y}$ and $\mathrm{La_{2-x}Sr_xCuO_4}$ as intrinsic left-handed materials. Because of the huge anisotropy, the negative refraction angle is small and the transmission coefficient for the backward wave is tiny at the interface between these cuprate superconductors and vacuum. The negative refraction angle and transmission coefficient can be improved in less anisotropic superconductors or using superconductor/dielectric superlattices. Recently it has been shown that anisotropic superconductors can emit a strong monochromatic THz waves \cite{ozyuzer_emission_2007,PhysRevLett.100.247006,SZLin_phase_2010}. The anisotropic superconductors can also be used as detectors and amplifiers \cite{PhysRevB.78.224519,PhysRevB.82.020504,dienst_optical_2013,rajasekaran_parametric_2016}. Combined with the left-handed propertied revealed here, it is expected that anisotropic superconductors can be used in novel electromagnetic devices with low loss.

\begin{acknowledgments}
We thank Alex Koshelev and Ulrich Welp for helpful discussions. This work was carried out under the auspices of the U.S. DOE contract No. DE-AC52-06NA25396 through the LDRD program.
\end{acknowledgments}

\bibliography{reference}

\begin{thebibliography}{38}%
\makeatletter
\providecommand \@ifxundefined [1]{%
 \@ifx{#1\undefined}
}%
\providecommand \@ifnum [1]{%
 \ifnum #1\expandafter \@firstoftwo
 \else \expandafter \@secondoftwo
 \fi
}%
\providecommand \@ifx [1]{%
 \ifx #1\expandafter \@firstoftwo
 \else \expandafter \@secondoftwo
 \fi
}%
\providecommand \natexlab [1]{#1}%
\providecommand \enquote  [1]{``#1''}%
\providecommand \bibnamefont  [1]{#1}%
\providecommand \bibfnamefont [1]{#1}%
\providecommand \citenamefont [1]{#1}%
\providecommand \href@noop [0]{\@secondoftwo}%
\providecommand \href [0]{\begingroup \@sanitize@url \@href}%
\providecommand \@href[1]{\@@startlink{#1}\@@href}%
\providecommand \@@href[1]{\endgroup#1\@@endlink}%
\providecommand \@sanitize@url [0]{\catcode `\\12\catcode `\$12\catcode
  `\&12\catcode `\#12\catcode `\^12\catcode `\_12\catcode `\%12\relax}%
\providecommand \@@startlink[1]{}%
\providecommand \@@endlink[0]{}%
\providecommand \url  [0]{\begingroup\@sanitize@url \@url }%
\providecommand \@url [1]{\endgroup\@href {#1}{\urlprefix }}%
\providecommand \urlprefix  [0]{URL }%
\providecommand \Eprint [0]{\href }%
\providecommand \doibase [0]{http://dx.doi.org/}%
\providecommand \selectlanguage [0]{\@gobble}%
\providecommand \bibinfo  [0]{\@secondoftwo}%
\providecommand \bibfield  [0]{\@secondoftwo}%
\providecommand \translation [1]{[#1]}%
\providecommand \BibitemOpen [0]{}%
\providecommand \bibitemStop [0]{}%
\providecommand \bibitemNoStop [0]{.\EOS\space}%
\providecommand \EOS [0]{\spacefactor3000\relax}%
\providecommand \BibitemShut  [1]{\csname bibitem#1\endcsname}%
\let\auto@bib@innerbib\@empty
\bibitem [{\citenamefont {Veselago}(1968)}]{Veselago1968}%
  \BibitemOpen
  \bibfield  {author} {\bibinfo {author} {\bibfnamefont {V.~G.}\ \bibnamefont
  {Veselago}},\ }\bibfield  {title} {\enquote {\bibinfo {title} {The
  electrodynamics of substances with simultaneously negative values of
  $\epsilon$ and $\mu$},}\ }\href {\doibase
  http://dx.doi.org/10.1070/PU1968v010n04ABEH003699} {\bibfield  {journal}
  {\bibinfo  {journal} {Sov. Phys. Uspekhi}\ }\textbf {\bibinfo {volume}
  {10}},\ \bibinfo {pages} {509} (\bibinfo {year} {1968})}\BibitemShut
  {NoStop}%
\bibitem [{\citenamefont {Shalaev}(2007)}]{shalaev_optical_2007}%
  \BibitemOpen
  \bibfield  {author} {\bibinfo {author} {\bibfnamefont {Vladimir~M.}\
  \bibnamefont {Shalaev}},\ }\bibfield  {title} {\enquote {\bibinfo {title}
  {Optical negative-index metamaterials},}\ }\href {\doibase
  10.1038/nphoton.2006.49} {\bibfield  {journal} {\bibinfo  {journal} {Nature
  Photonics}\ }\textbf {\bibinfo {volume} {1}},\ \bibinfo {pages} {41--48}
  (\bibinfo {year} {2007})}\BibitemShut {NoStop}%
\bibitem [{\citenamefont {Chen}\ \emph {et~al.}(2016)\citenamefont {Chen},
  \citenamefont {Taylor},\ and\ \citenamefont {Yu}}]{chen_review_2016}%
  \BibitemOpen
  \bibfield  {author} {\bibinfo {author} {\bibfnamefont {Hou-Tong}\
  \bibnamefont {Chen}}, \bibinfo {author} {\bibfnamefont {Antoinette~J.}\
  \bibnamefont {Taylor}}, \ and\ \bibinfo {author} {\bibfnamefont {Nanfang}\
  \bibnamefont {Yu}},\ }\bibfield  {title} {\enquote {\bibinfo {title} {A
  review of metasurfaces: physics and applications},}\ }\href {\doibase
  10.1088/0034-4885/79/7/076401} {\bibfield  {journal} {\bibinfo  {journal}
  {Reports on Progress in Physics}\ }\textbf {\bibinfo {volume} {79}},\
  \bibinfo {pages} {076401} (\bibinfo {year} {2016})}\BibitemShut {NoStop}%
\bibitem [{\citenamefont {Smith}\ \emph {et~al.}(2000)\citenamefont {Smith},
  \citenamefont {Padilla}, \citenamefont {Vier}, \citenamefont {Nemat-Nasser},\
  and\ \citenamefont {Schultz}}]{PhysRevLett.84.4184}%
  \BibitemOpen
  \bibfield  {author} {\bibinfo {author} {\bibfnamefont {D.~R.}\ \bibnamefont
  {Smith}}, \bibinfo {author} {\bibfnamefont {Willie~J.}\ \bibnamefont
  {Padilla}}, \bibinfo {author} {\bibfnamefont {D.~C.}\ \bibnamefont {Vier}},
  \bibinfo {author} {\bibfnamefont {S.~C.}\ \bibnamefont {Nemat-Nasser}}, \
  and\ \bibinfo {author} {\bibfnamefont {S.}~\bibnamefont {Schultz}},\
  }\bibfield  {title} {\enquote {\bibinfo {title} {Composite medium with
  simultaneously negative permeability and permittivity},}\ }\href {\doibase
  10.1103/PhysRevLett.84.4184} {\bibfield  {journal} {\bibinfo  {journal}
  {Phys. Rev. Lett.}\ }\textbf {\bibinfo {volume} {84}},\ \bibinfo {pages}
  {4184--4187} (\bibinfo {year} {2000})}\BibitemShut {NoStop}%
\bibitem [{\citenamefont {Shelby}\ \emph {et~al.}(2001)\citenamefont {Shelby},
  \citenamefont {Smith},\ and\ \citenamefont
  {Schultz}}]{shelby_experimental_2001}%
  \BibitemOpen
  \bibfield  {author} {\bibinfo {author} {\bibfnamefont {R.~A.}\ \bibnamefont
  {Shelby}}, \bibinfo {author} {\bibfnamefont {D.~R.}\ \bibnamefont {Smith}}, \
  and\ \bibinfo {author} {\bibfnamefont {S.}~\bibnamefont {Schultz}},\
  }\bibfield  {title} {\enquote {\bibinfo {title} {Experimental {Verification}
  of a {Negative} {Index} of {Refraction}},}\ }\href {\doibase
  10.1126/science.1058847} {\bibfield  {journal} {\bibinfo  {journal}
  {Science}\ }\textbf {\bibinfo {volume} {292}},\ \bibinfo {pages} {77--79}
  (\bibinfo {year} {2001})}\BibitemShut {NoStop}%
\bibitem [{\citenamefont {Smith}\ and\ \citenamefont
  {Schurig}(2003)}]{PhysRevLett.90.077405}%
  \BibitemOpen
  \bibfield  {author} {\bibinfo {author} {\bibfnamefont {D.~R.}\ \bibnamefont
  {Smith}}\ and\ \bibinfo {author} {\bibfnamefont {D.}~\bibnamefont
  {Schurig}},\ }\bibfield  {title} {\enquote {\bibinfo {title} {Electromagnetic
  wave propagation in media with indefinite permittivity and permeability
  tensors},}\ }\href {\doibase 10.1103/PhysRevLett.90.077405} {\bibfield
  {journal} {\bibinfo  {journal} {Phys. Rev. Lett.}\ }\textbf {\bibinfo
  {volume} {90}},\ \bibinfo {pages} {077405} (\bibinfo {year}
  {2003})}\BibitemShut {NoStop}%
\bibitem [{\citenamefont {Belov}(2003)}]{belov_backward_2003}%
  \BibitemOpen
  \bibfield  {author} {\bibinfo {author} {\bibfnamefont {P.~A.}\ \bibnamefont
  {Belov}},\ }\bibfield  {title} {\enquote {\bibinfo {title} {Backward waves
  and negative refraction in uniaxial dielectrics with negative dielectric
  permittivity along the anisotropy axis},}\ }\href {\doibase
  10.1002/mop.10887} {\bibfield  {journal} {\bibinfo  {journal} {Microwave and
  Optical Technology Letters}\ }\textbf {\bibinfo {volume} {37}},\ \bibinfo
  {pages} {259--263} (\bibinfo {year} {2003})}\BibitemShut {NoStop}%
\bibitem [{\citenamefont {Poddubny}\ \emph {et~al.}(2013)\citenamefont
  {Poddubny}, \citenamefont {Iorsh}, \citenamefont {Belov},\ and\ \citenamefont
  {Kivshar}}]{poddubny_hyperbolic_2013}%
  \BibitemOpen
  \bibfield  {author} {\bibinfo {author} {\bibfnamefont {Alexander}\
  \bibnamefont {Poddubny}}, \bibinfo {author} {\bibfnamefont {Ivan}\
  \bibnamefont {Iorsh}}, \bibinfo {author} {\bibfnamefont {Pavel}\ \bibnamefont
  {Belov}}, \ and\ \bibinfo {author} {\bibfnamefont {Yuri}\ \bibnamefont
  {Kivshar}},\ }\bibfield  {title} {\enquote {\bibinfo {title} {Hyperbolic
  metamaterials},}\ }\href {\doibase 10.1038/nphoton.2013.243} {\bibfield
  {journal} {\bibinfo  {journal} {Nature Photonics}\ }\textbf {\bibinfo
  {volume} {7}},\ \bibinfo {pages} {948--957} (\bibinfo {year}
  {2013})}\BibitemShut {NoStop}%
\bibitem [{\citenamefont {Cortes}\ \emph {et~al.}(2012)\citenamefont {Cortes},
  \citenamefont {Newman}, \citenamefont {Molesky},\ and\ \citenamefont
  {Jacob}}]{cortes_quantum_2012}%
  \BibitemOpen
  \bibfield  {author} {\bibinfo {author} {\bibfnamefont {C.~L.}\ \bibnamefont
  {Cortes}}, \bibinfo {author} {\bibfnamefont {W.}~\bibnamefont {Newman}},
  \bibinfo {author} {\bibfnamefont {S.}~\bibnamefont {Molesky}}, \ and\
  \bibinfo {author} {\bibfnamefont {Z.}~\bibnamefont {Jacob}},\ }\bibfield
  {title} {\enquote {\bibinfo {title} {Quantum nanophotonics using hyperbolic
  metamaterials},}\ }\href {\doibase 10.1088/2040-8978/14/6/063001} {\bibfield
  {journal} {\bibinfo  {journal} {Journal of Optics}\ }\textbf {\bibinfo
  {volume} {14}},\ \bibinfo {pages} {063001} (\bibinfo {year}
  {2012})}\BibitemShut {NoStop}%
\bibitem [{\citenamefont {Agranovich}\ and\ \citenamefont
  {Kravtsov}(1985)}]{agranovich_notes_1985}%
  \BibitemOpen
  \bibfield  {author} {\bibinfo {author} {\bibfnamefont {V.~M.}\ \bibnamefont
  {Agranovich}}\ and\ \bibinfo {author} {\bibfnamefont {V.~E.}\ \bibnamefont
  {Kravtsov}},\ }\bibfield  {title} {\enquote {\bibinfo {title} {Notes on
  crystal optics of superlattices},}\ }\href {\doibase
  10.1016/0038-1098(85)91111-1} {\bibfield  {journal} {\bibinfo  {journal}
  {Solid State Communications}\ }\textbf {\bibinfo {volume} {55}},\ \bibinfo
  {pages} {85--90} (\bibinfo {year} {1985})}\BibitemShut {NoStop}%
\bibitem [{\citenamefont {Yoxall}\ \emph {et~al.}(2015)\citenamefont {Yoxall},
  \citenamefont {Schnell}, \citenamefont {Nikitin}, \citenamefont {Txoperena},
  \citenamefont {Woessner}, \citenamefont {Lundeberg}, \citenamefont
  {Casanova}, \citenamefont {Hueso}, \citenamefont {Koppens},\ and\
  \citenamefont {Hillenbrand}}]{yoxall_direct_2015}%
  \BibitemOpen
  \bibfield  {author} {\bibinfo {author} {\bibfnamefont {Edward}\ \bibnamefont
  {Yoxall}}, \bibinfo {author} {\bibfnamefont {Martin}\ \bibnamefont
  {Schnell}}, \bibinfo {author} {\bibfnamefont {Alexey~Y.}\ \bibnamefont
  {Nikitin}}, \bibinfo {author} {\bibfnamefont {Oihana}\ \bibnamefont
  {Txoperena}}, \bibinfo {author} {\bibfnamefont {Achim}\ \bibnamefont
  {Woessner}}, \bibinfo {author} {\bibfnamefont {Mark~B.}\ \bibnamefont
  {Lundeberg}}, \bibinfo {author} {\bibfnamefont {Félix}\ \bibnamefont
  {Casanova}}, \bibinfo {author} {\bibfnamefont {Luis~E.}\ \bibnamefont
  {Hueso}}, \bibinfo {author} {\bibfnamefont {Frank H.~L.}\ \bibnamefont
  {Koppens}}, \ and\ \bibinfo {author} {\bibfnamefont {Rainer}\ \bibnamefont
  {Hillenbrand}},\ }\bibfield  {title} {\enquote {\bibinfo {title} {Direct
  observation of ultraslow hyperbolic polariton propagation with negative phase
  velocity},}\ }\href {\doibase 10.1038/nphoton.2015.166} {\bibfield  {journal}
  {\bibinfo  {journal} {Nature Photonics}\ }\textbf {\bibinfo {volume} {9}},\
  \bibinfo {pages} {674--678} (\bibinfo {year} {2015})}\BibitemShut {NoStop}%
\bibitem [{\citenamefont {Kleiner}\ \emph {et~al.}(1992)\citenamefont
  {Kleiner}, \citenamefont {Steinmeyer}, \citenamefont {Kunkel},\ and\
  \citenamefont {M\"uller}}]{Kleiner92}%
  \BibitemOpen
  \bibfield  {author} {\bibinfo {author} {\bibfnamefont {R.}~\bibnamefont
  {Kleiner}}, \bibinfo {author} {\bibfnamefont {F.}~\bibnamefont {Steinmeyer}},
  \bibinfo {author} {\bibfnamefont {G.}~\bibnamefont {Kunkel}}, \ and\ \bibinfo
  {author} {\bibfnamefont {P.}~\bibnamefont {M\"uller}},\ }\bibfield  {title}
  {\enquote {\bibinfo {title} {Intrinsic josephson effects in
  $\mathrm{Bi_2Sr_2CaCu_2O_{8+\delta}}$ single crystals},}\ }\href {\doibase
  10.1103/PhysRevLett.68.2394} {\bibfield  {journal} {\bibinfo  {journal}
  {Phys. Rev. Lett.}\ }\textbf {\bibinfo {volume} {68}},\ \bibinfo {pages}
  {2394--2397} (\bibinfo {year} {1992})}\BibitemShut {NoStop}%
\bibitem [{\citenamefont {Kleiner}\ and\ \citenamefont
  {M\"uller}(1994)}]{PhysRevB.49.1327}%
  \BibitemOpen
  \bibfield  {author} {\bibinfo {author} {\bibfnamefont {R.}~\bibnamefont
  {Kleiner}}\ and\ \bibinfo {author} {\bibfnamefont {P.}~\bibnamefont
  {M\"uller}},\ }\bibfield  {title} {\enquote {\bibinfo {title} {Intrinsic
  josephson effects in high-$\mathrm{T}_c$ superconductors},}\ }\href {\doibase
  10.1103/PhysRevB.49.1327} {\bibfield  {journal} {\bibinfo  {journal} {Phys.
  Rev. B}\ }\textbf {\bibinfo {volume} {49}},\ \bibinfo {pages} {1327--1341}
  (\bibinfo {year} {1994})}\BibitemShut {NoStop}%
\bibitem [{\citenamefont {Tachiki}\ \emph {et~al.}(1994)\citenamefont
  {Tachiki}, \citenamefont {Koyama},\ and\ \citenamefont
  {Takahashi}}]{Tachiki94}%
  \BibitemOpen
  \bibfield  {author} {\bibinfo {author} {\bibfnamefont {M.}~\bibnamefont
  {Tachiki}}, \bibinfo {author} {\bibfnamefont {T.}~\bibnamefont {Koyama}}, \
  and\ \bibinfo {author} {\bibfnamefont {S.}~\bibnamefont {Takahashi}},\
  }\bibfield  {title} {\enquote {\bibinfo {title} {Electromagnetic phenomena
  related to a low-frequency plasma in cuprate superconductors},}\ }\href
  {\doibase 10.1103/PhysRevB.50.7065} {\bibfield  {journal} {\bibinfo
  {journal} {Phys. Rev. B}\ }\textbf {\bibinfo {volume} {50}},\ \bibinfo
  {pages} {7065--7084} (\bibinfo {year} {1994})}\BibitemShut {NoStop}%
\bibitem [{\citenamefont {Bulaevskii}\ \emph {et~al.}(1994)\citenamefont
  {Bulaevskii}, \citenamefont {Zamora}, \citenamefont {Baeriswyl},
  \citenamefont {Beck},\ and\ \citenamefont {Clem}}]{Bulaevskii94}%
  \BibitemOpen
  \bibfield  {author} {\bibinfo {author} {\bibfnamefont {L.~N.}\ \bibnamefont
  {Bulaevskii}}, \bibinfo {author} {\bibfnamefont {M.}~\bibnamefont {Zamora}},
  \bibinfo {author} {\bibfnamefont {D.}~\bibnamefont {Baeriswyl}}, \bibinfo
  {author} {\bibfnamefont {H.}~\bibnamefont {Beck}}, \ and\ \bibinfo {author}
  {\bibfnamefont {John~R.}\ \bibnamefont {Clem}},\ }\bibfield  {title}
  {\enquote {\bibinfo {title} {Time-dependent equations for phase differences
  and a collective mode in josephson-coupled layered superconductors},}\ }\href
  {\doibase 10.1103/PhysRevB.50.12831} {\bibfield  {journal} {\bibinfo
  {journal} {Phys. Rev. B}\ }\textbf {\bibinfo {volume} {50}},\ \bibinfo
  {pages} {12831--12834} (\bibinfo {year} {1994})}\BibitemShut {NoStop}%
\bibitem [{\citenamefont {Ferguson}\ and\ \citenamefont
  {Zhang}(2002)}]{Ferguson02}%
  \BibitemOpen
  \bibfield  {author} {\bibinfo {author} {\bibfnamefont {Bradley}\ \bibnamefont
  {Ferguson}}\ and\ \bibinfo {author} {\bibfnamefont {Xi-Cheng}\ \bibnamefont
  {Zhang}},\ }\bibfield  {title} {\enquote {\bibinfo {title} {Materials for
  terahertz science and technology},}\ }\href {\doibase 10.1038/nmat708}
  {\bibfield  {journal} {\bibinfo  {journal} {Nature Materials}\ }\textbf
  {\bibinfo {volume} {1}},\ \bibinfo {pages} {26--33} (\bibinfo {year}
  {2002})}\BibitemShut {NoStop}%
\bibitem [{\citenamefont {Tonouchi}(2007)}]{Tonouchi07}%
  \BibitemOpen
  \bibfield  {author} {\bibinfo {author} {\bibfnamefont {Masayoshi}\
  \bibnamefont {Tonouchi}},\ }\bibfield  {title} {\enquote {\bibinfo {title}
  {Cutting-edge terahertz technology},}\ }\href {\doibase
  10.1038/nphoton.2007.3} {\bibfield  {journal} {\bibinfo  {journal} {Nature
  Photonics}\ }\textbf {\bibinfo {volume} {1}},\ \bibinfo {pages} {97--105}
  (\bibinfo {year} {2007})}\BibitemShut {NoStop}%
\bibitem [{\citenamefont {Basov}\ \emph {et~al.}(1994)\citenamefont {Basov},
  \citenamefont {Timusk}, \citenamefont {Dabrowski},\ and\ \citenamefont
  {Jorgensen}}]{PhysRevB.50.3511}%
  \BibitemOpen
  \bibfield  {author} {\bibinfo {author} {\bibfnamefont {D.~N.}\ \bibnamefont
  {Basov}}, \bibinfo {author} {\bibfnamefont {T.}~\bibnamefont {Timusk}},
  \bibinfo {author} {\bibfnamefont {B.}~\bibnamefont {Dabrowski}}, \ and\
  \bibinfo {author} {\bibfnamefont {J.~D.}\ \bibnamefont {Jorgensen}},\
  }\bibfield  {title} {\enquote {\bibinfo {title} {\textit{c} -axis response of
  $\mathrm{YBa_2Cu_4O_{8}}$: A pseudogap and possibility of josephson coupling
  of $\mathrm{CuO_2}$ planes},}\ }\href {\doibase 10.1103/PhysRevB.50.3511}
  {\bibfield  {journal} {\bibinfo  {journal} {Phys. Rev. B}\ }\textbf {\bibinfo
  {volume} {50}},\ \bibinfo {pages} {3511--3514} (\bibinfo {year}
  {1994})}\BibitemShut {NoStop}%
\bibitem [{\citenamefont {Kojima}\ \emph {et~al.}(2002)\citenamefont {Kojima},
  \citenamefont {Uchida}, \citenamefont {Fudamoto},\ and\ \citenamefont
  {Tajima}}]{PhysRevLett.89.247001}%
  \BibitemOpen
  \bibfield  {author} {\bibinfo {author} {\bibfnamefont {K.~M.}\ \bibnamefont
  {Kojima}}, \bibinfo {author} {\bibfnamefont {S.}~\bibnamefont {Uchida}},
  \bibinfo {author} {\bibfnamefont {Y.}~\bibnamefont {Fudamoto}}, \ and\
  \bibinfo {author} {\bibfnamefont {S.}~\bibnamefont {Tajima}},\ }\bibfield
  {title} {\enquote {\bibinfo {title} {New josephson plasma modes in underdoped
  $\mathrm{YBa_2Cu_3O_{6.6}}$ induced by a parallel magnetic field},}\ }\href
  {\doibase 10.1103/PhysRevLett.89.247001} {\bibfield  {journal} {\bibinfo
  {journal} {Phys. Rev. Lett.}\ }\textbf {\bibinfo {volume} {89}},\ \bibinfo
  {pages} {247001} (\bibinfo {year} {2002})}\BibitemShut {NoStop}%
\bibitem [{\citenamefont {Tamasaku}\ \emph {et~al.}(1992)\citenamefont
  {Tamasaku}, \citenamefont {Nakamura},\ and\ \citenamefont
  {Uchida}}]{PhysRevLett.69.1455}%
  \BibitemOpen
  \bibfield  {author} {\bibinfo {author} {\bibfnamefont {K.}~\bibnamefont
  {Tamasaku}}, \bibinfo {author} {\bibfnamefont {Y.}~\bibnamefont {Nakamura}},
  \ and\ \bibinfo {author} {\bibfnamefont {S.}~\bibnamefont {Uchida}},\
  }\bibfield  {title} {\enquote {\bibinfo {title} {Charge dynamics across the
  $\mathrm{CuO_2}$ planes in $\mathrm{La_{2-x}Sr_xCuO_{4}}$},}\ }\href
  {\doibase 10.1103/PhysRevLett.69.1455} {\bibfield  {journal} {\bibinfo
  {journal} {Phys. Rev. Lett.}\ }\textbf {\bibinfo {volume} {69}},\ \bibinfo
  {pages} {1455--1458} (\bibinfo {year} {1992})}\BibitemShut {NoStop}%
\bibitem [{\citenamefont {Shibauchi}\ \emph {et~al.}(1994)\citenamefont
  {Shibauchi}, \citenamefont {Kitano}, \citenamefont {Uchinokura},
  \citenamefont {Maeda}, \citenamefont {Kimura},\ and\ \citenamefont
  {Kishio}}]{PhysRevLett.72.2263}%
  \BibitemOpen
  \bibfield  {author} {\bibinfo {author} {\bibfnamefont {T.}~\bibnamefont
  {Shibauchi}}, \bibinfo {author} {\bibfnamefont {H.}~\bibnamefont {Kitano}},
  \bibinfo {author} {\bibfnamefont {K.}~\bibnamefont {Uchinokura}}, \bibinfo
  {author} {\bibfnamefont {A.}~\bibnamefont {Maeda}}, \bibinfo {author}
  {\bibfnamefont {T.}~\bibnamefont {Kimura}}, \ and\ \bibinfo {author}
  {\bibfnamefont {K.}~\bibnamefont {Kishio}},\ }\bibfield  {title} {\enquote
  {\bibinfo {title} {Anisotropic penetration depth in $\mathrm{CuO_2}$ planes
  in $\mathrm{La_{2-x}Sr_xCuO_{4}}$},}\ }\href {\doibase
  10.1103/PhysRevLett.72.2263} {\bibfield  {journal} {\bibinfo  {journal}
  {Phys. Rev. Lett.}\ }\textbf {\bibinfo {volume} {72}},\ \bibinfo {pages}
  {2263--2266} (\bibinfo {year} {1994})}\BibitemShut {NoStop}%
\bibitem [{\citenamefont {Shibauchi}\ \emph {et~al.}(1997)\citenamefont
  {Shibauchi}, \citenamefont {Sato}, \citenamefont {Mashio}, \citenamefont
  {Tamegai}, \citenamefont {Mori}, \citenamefont {Tajima},\ and\ \citenamefont
  {Tanaka}}]{PhysRevB.55.R11977}%
  \BibitemOpen
  \bibfield  {author} {\bibinfo {author} {\bibfnamefont {T.}~\bibnamefont
  {Shibauchi}}, \bibinfo {author} {\bibfnamefont {M.}~\bibnamefont {Sato}},
  \bibinfo {author} {\bibfnamefont {A.}~\bibnamefont {Mashio}}, \bibinfo
  {author} {\bibfnamefont {T.}~\bibnamefont {Tamegai}}, \bibinfo {author}
  {\bibfnamefont {H.}~\bibnamefont {Mori}}, \bibinfo {author} {\bibfnamefont
  {S.}~\bibnamefont {Tajima}}, \ and\ \bibinfo {author} {\bibfnamefont
  {S.}~\bibnamefont {Tanaka}},\ }\bibfield  {title} {\enquote {\bibinfo {title}
  {Josephson plasma resonance in the mixed state of the organic superconductor
  \ensuremath{\kappa}-(bedt-ttf${)}_{2}$cu(ncs${)}_{2}$},}\ }\href {\doibase
  10.1103/PhysRevB.55.R11977} {\bibfield  {journal} {\bibinfo  {journal} {Phys.
  Rev. B}\ }\textbf {\bibinfo {volume} {55}},\ \bibinfo {pages}
  {R11977--R11980} (\bibinfo {year} {1997})}\BibitemShut {NoStop}%
\bibitem [{\citenamefont {Mola}\ \emph {et~al.}(2000)\citenamefont {Mola},
  \citenamefont {King}, \citenamefont {McRaven}, \citenamefont {Hill},
  \citenamefont {Qualls},\ and\ \citenamefont {Brooks}}]{PhysRevB.62.5965}%
  \BibitemOpen
  \bibfield  {author} {\bibinfo {author} {\bibfnamefont {M.~M.}\ \bibnamefont
  {Mola}}, \bibinfo {author} {\bibfnamefont {J.~T.}\ \bibnamefont {King}},
  \bibinfo {author} {\bibfnamefont {C.~P.}\ \bibnamefont {McRaven}}, \bibinfo
  {author} {\bibfnamefont {S.}~\bibnamefont {Hill}}, \bibinfo {author}
  {\bibfnamefont {J.~S.}\ \bibnamefont {Qualls}}, \ and\ \bibinfo {author}
  {\bibfnamefont {J.~S.}\ \bibnamefont {Brooks}},\ }\bibfield  {title}
  {\enquote {\bibinfo {title} {Josephson plasma resonance in
  $\ensuremath{\kappa}-(\mathrm{BEDT}-\mathrm{TTF}{)}_{2}\mathrm{Cu}(\mathrm{NCS}{)}_{2}$},}\
  }\href {\doibase 10.1103/PhysRevB.62.5965} {\bibfield  {journal} {\bibinfo
  {journal} {Phys. Rev. B}\ }\textbf {\bibinfo {volume} {62}},\ \bibinfo
  {pages} {5965--5970} (\bibinfo {year} {2000})}\BibitemShut {NoStop}%
\bibitem [{\citenamefont {Koshelev}\ and\ \citenamefont
  {Aranson}(2001)}]{Koshelev01}%
  \BibitemOpen
  \bibfield  {author} {\bibinfo {author} {\bibfnamefont {A.~E.}\ \bibnamefont
  {Koshelev}}\ and\ \bibinfo {author} {\bibfnamefont {I.}~\bibnamefont
  {Aranson}},\ }\bibfield  {title} {\enquote {\bibinfo {title} {Dynamic
  structure selection and instabilities of driven josephson lattice in
  high-temperature superconductors},}\ }\href {\doibase
  10.1103/PhysRevB.64.174508} {\bibfield  {journal} {\bibinfo  {journal} {Phys.
  Rev. B}\ }\textbf {\bibinfo {volume} {64}},\ \bibinfo {pages} {174508}
  (\bibinfo {year} {2001})}\BibitemShut {NoStop}%
\bibitem [{\citenamefont {Latyshev}\ \emph {et~al.}(1999)\citenamefont
  {Latyshev}, \citenamefont {Yamashita}, \citenamefont {Bulaevskii},
  \citenamefont {Graf}, \citenamefont {Balatsky},\ and\ \citenamefont
  {Maley}}]{Latyshev1999}%
  \BibitemOpen
  \bibfield  {author} {\bibinfo {author} {\bibfnamefont {Yu.~I.}\ \bibnamefont
  {Latyshev}}, \bibinfo {author} {\bibfnamefont {T.}~\bibnamefont {Yamashita}},
  \bibinfo {author} {\bibfnamefont {L.~N.}\ \bibnamefont {Bulaevskii}},
  \bibinfo {author} {\bibfnamefont {M.~J.}\ \bibnamefont {Graf}}, \bibinfo
  {author} {\bibfnamefont {A.~V.}\ \bibnamefont {Balatsky}}, \ and\ \bibinfo
  {author} {\bibfnamefont {M.~P.}\ \bibnamefont {Maley}},\ }\bibfield  {title}
  {\enquote {\bibinfo {title} {Interlayer transport of quasiparticles and
  cooper pairs in $\mathrm{Bi_2Sr_2CaCu_2O_{8+\delta}}$ superconductors},}\
  }\href {\doibase 10.1103/PhysRevLett.82.5345} {\bibfield  {journal} {\bibinfo
   {journal} {Phys. Rev. Lett.}\ }\textbf {\bibinfo {volume} {82}},\ \bibinfo
  {pages} {5345--5348} (\bibinfo {year} {1999})}\BibitemShut {NoStop}%
\bibitem [{\citenamefont {Corson}\ \emph {et~al.}(2000)\citenamefont {Corson},
  \citenamefont {Orenstein}, \citenamefont {Oh}, \citenamefont {O'Donnell},\
  and\ \citenamefont {Eckstein}}]{Corson2000}%
  \BibitemOpen
  \bibfield  {author} {\bibinfo {author} {\bibfnamefont {J.}~\bibnamefont
  {Corson}}, \bibinfo {author} {\bibfnamefont {J.}~\bibnamefont {Orenstein}},
  \bibinfo {author} {\bibfnamefont {Seongshik}\ \bibnamefont {Oh}}, \bibinfo
  {author} {\bibfnamefont {J.}~\bibnamefont {O'Donnell}}, \ and\ \bibinfo
  {author} {\bibfnamefont {J.~N.}\ \bibnamefont {Eckstein}},\ }\bibfield
  {title} {\enquote {\bibinfo {title} {Nodal quasiparticle lifetime in the
  superconducting state of $\mathrm{Bi_2Sr_2CaCu_2O_{8+\delta}}$},}\ }\href
  {\doibase 10.1103/PhysRevLett.85.2569} {\bibfield  {journal} {\bibinfo
  {journal} {Phys. Rev. Lett.}\ }\textbf {\bibinfo {volume} {85}},\ \bibinfo
  {pages} {2569--2572} (\bibinfo {year} {2000})}\BibitemShut {NoStop}%
\bibitem [{\citenamefont {Grischkowsky}\ \emph {et~al.}(1990)\citenamefont
  {Grischkowsky}, \citenamefont {Keiding}, \citenamefont {van Exter},\ and\
  \citenamefont {Fattinger}}]{Grischkowsky90}%
  \BibitemOpen
  \bibfield  {author} {\bibinfo {author} {\bibfnamefont {D.}~\bibnamefont
  {Grischkowsky}}, \bibinfo {author} {\bibfnamefont {S{\o}ren}\ \bibnamefont
  {Keiding}}, \bibinfo {author} {\bibfnamefont {Martin}\ \bibnamefont {van
  Exter}}, \ and\ \bibinfo {author} {\bibfnamefont {Ch.}\ \bibnamefont
  {Fattinger}},\ }\bibfield  {title} {\enquote {\bibinfo {title} {Far-infrared
  time-domain spectroscopy with terahertz beams of dielectrics and
  semiconductors},}\ }\href {\doibase 10.1364/JOSAB.7.002006} {\bibfield
  {journal} {\bibinfo  {journal} {J. Opt. Soc. Am. B}\ }\textbf {\bibinfo
  {volume} {7}},\ \bibinfo {pages} {2006--2015} (\bibinfo {year}
  {1990})}\BibitemShut {NoStop}%
\bibitem [{\citenamefont {Zitzmann}\ \emph {et~al.}(2002)\citenamefont
  {Zitzmann}, \citenamefont {Ustinov}, \citenamefont {Levitchev},\ and\
  \citenamefont {Sakai}}]{PhysRevB.66.064527}%
  \BibitemOpen
  \bibfield  {author} {\bibinfo {author} {\bibfnamefont {J.}~\bibnamefont
  {Zitzmann}}, \bibinfo {author} {\bibfnamefont {A.~V.}\ \bibnamefont
  {Ustinov}}, \bibinfo {author} {\bibfnamefont {M.}~\bibnamefont {Levitchev}},
  \ and\ \bibinfo {author} {\bibfnamefont {S.}~\bibnamefont {Sakai}},\
  }\bibfield  {title} {\enquote {\bibinfo {title} {Super-relativistic fluxon in
  a josephson multilayer: Experiment and simulation},}\ }\href {\doibase
  10.1103/PhysRevB.66.064527} {\bibfield  {journal} {\bibinfo  {journal} {Phys.
  Rev. B}\ }\textbf {\bibinfo {volume} {66}},\ \bibinfo {pages} {064527}
  (\bibinfo {year} {2002})}\BibitemShut {NoStop}%
\bibitem [{\citenamefont {Goldobin}\ \emph {et~al.}(2000)\citenamefont
  {Goldobin}, \citenamefont {Wallraff},\ and\ \citenamefont
  {Ustinov}}]{goldobin_cherenkov_2000}%
  \BibitemOpen
  \bibfield  {author} {\bibinfo {author} {\bibfnamefont {E.}~\bibnamefont
  {Goldobin}}, \bibinfo {author} {\bibfnamefont {A.}~\bibnamefont {Wallraff}},
  \ and\ \bibinfo {author} {\bibfnamefont {A.~V.}\ \bibnamefont {Ustinov}},\
  }\bibfield  {title} {\enquote {\bibinfo {title} {Cherenkov {Radiation} from
  {Fluxon} in a {Stack} of {Coupled} {Long} {Josephson} {Junctions}},}\ }\href
  {\doibase 10.1023/A:1004677528120} {\bibfield  {journal} {\bibinfo  {journal}
  {Journal of Low Temperature Physics}\ }\textbf {\bibinfo {volume} {119}},\
  \bibinfo {pages} {589--614} (\bibinfo {year} {2000})}\BibitemShut {NoStop}%
\bibitem [{\citenamefont {Sakai}\ \emph {et~al.}(1993)\citenamefont {Sakai},
  \citenamefont {Bodin},\ and\ \citenamefont {Pedersen}}]{sakai_fluxons_1993}%
  \BibitemOpen
  \bibfield  {author} {\bibinfo {author} {\bibfnamefont {S.}~\bibnamefont
  {Sakai}}, \bibinfo {author} {\bibfnamefont {P.}~\bibnamefont {Bodin}}, \ and\
  \bibinfo {author} {\bibfnamefont {N.~F.}\ \bibnamefont {Pedersen}},\
  }\bibfield  {title} {\enquote {\bibinfo {title} {Fluxons in thin‐film
  superconductor‐insulator superlattices},}\ }\href {\doibase
  10.1063/1.353095} {\bibfield  {journal} {\bibinfo  {journal} {Journal of
  Applied Physics}\ }\textbf {\bibinfo {volume} {73}},\ \bibinfo {pages}
  {2411--2418} (\bibinfo {year} {1993})}\BibitemShut {NoStop}%
\bibitem [{\citenamefont {Kleiner}(1994)}]{PhysRevB.50.6919}%
  \BibitemOpen
  \bibfield  {author} {\bibinfo {author} {\bibfnamefont {R.}~\bibnamefont
  {Kleiner}},\ }\bibfield  {title} {\enquote {\bibinfo {title} {Two-dimensional
  resonant modes in stacked josephson junctions},}\ }\href {\doibase
  10.1103/PhysRevB.50.6919} {\bibfield  {journal} {\bibinfo  {journal} {Phys.
  Rev. B}\ }\textbf {\bibinfo {volume} {50}},\ \bibinfo {pages} {6919--6922}
  (\bibinfo {year} {1994})}\BibitemShut {NoStop}%
\bibitem [{\citenamefont {Ozyuzer}\ \emph {et~al.}(2007)\citenamefont
  {Ozyuzer}, \citenamefont {Koshelev}, \citenamefont {Kurter}, \citenamefont
  {Gopalsami}, \citenamefont {Li}, \citenamefont {Tachiki}, \citenamefont
  {Kadowaki}, \citenamefont {Yamamoto}, \citenamefont {Minami}, \citenamefont
  {Yamaguchi}, \citenamefont {Tachiki}, \citenamefont {Gray}, \citenamefont
  {Kwok},\ and\ \citenamefont {Welp}}]{ozyuzer_emission_2007}%
  \BibitemOpen
  \bibfield  {author} {\bibinfo {author} {\bibfnamefont {L.}~\bibnamefont
  {Ozyuzer}}, \bibinfo {author} {\bibfnamefont {A.~E.}\ \bibnamefont
  {Koshelev}}, \bibinfo {author} {\bibfnamefont {C.}~\bibnamefont {Kurter}},
  \bibinfo {author} {\bibfnamefont {N.}~\bibnamefont {Gopalsami}}, \bibinfo
  {author} {\bibfnamefont {Q.}~\bibnamefont {Li}}, \bibinfo {author}
  {\bibfnamefont {M.}~\bibnamefont {Tachiki}}, \bibinfo {author} {\bibfnamefont
  {K.}~\bibnamefont {Kadowaki}}, \bibinfo {author} {\bibfnamefont
  {T.}~\bibnamefont {Yamamoto}}, \bibinfo {author} {\bibfnamefont
  {H.}~\bibnamefont {Minami}}, \bibinfo {author} {\bibfnamefont
  {H.}~\bibnamefont {Yamaguchi}}, \bibinfo {author} {\bibfnamefont
  {T.}~\bibnamefont {Tachiki}}, \bibinfo {author} {\bibfnamefont {K.~E.}\
  \bibnamefont {Gray}}, \bibinfo {author} {\bibfnamefont {W.-K.}\ \bibnamefont
  {Kwok}}, \ and\ \bibinfo {author} {\bibfnamefont {U.}~\bibnamefont {Welp}},\
  }\bibfield  {title} {\enquote {\bibinfo {title} {Emission of {Coherent} {THz}
  {Radiation} from {Superconductors}},}\ }\href {\doibase
  10.1126/science.1149802} {\bibfield  {journal} {\bibinfo  {journal}
  {Science}\ }\textbf {\bibinfo {volume} {318}},\ \bibinfo {pages} {1291--1293}
  (\bibinfo {year} {2007})}\BibitemShut {NoStop}%
\bibitem [{\citenamefont {Lin}\ and\ \citenamefont
  {Hu}(2008)}]{PhysRevLett.100.247006}%
  \BibitemOpen
  \bibfield  {author} {\bibinfo {author} {\bibfnamefont {Shizeng}\ \bibnamefont
  {Lin}}\ and\ \bibinfo {author} {\bibfnamefont {Xiao}\ \bibnamefont {Hu}},\
  }\bibfield  {title} {\enquote {\bibinfo {title} {Possible dynamic states in
  inductively coupled intrinsic josephson junctions of layered high-${T}_{c}$
  superconductors},}\ }\href {\doibase 10.1103/PhysRevLett.100.247006}
  {\bibfield  {journal} {\bibinfo  {journal} {Phys. Rev. Lett.}\ }\textbf
  {\bibinfo {volume} {100}},\ \bibinfo {pages} {247006} (\bibinfo {year}
  {2008})}\BibitemShut {NoStop}%
\bibitem [{\citenamefont {Hu}\ and\ \citenamefont
  {Lin}(2010)}]{SZLin_phase_2010}%
  \BibitemOpen
  \bibfield  {author} {\bibinfo {author} {\bibfnamefont {Xiao}\ \bibnamefont
  {Hu}}\ and\ \bibinfo {author} {\bibfnamefont {Shi-Zeng}\ \bibnamefont
  {Lin}},\ }\bibfield  {title} {\enquote {\bibinfo {title} {Phase dynamics in a
  stack of inductively coupled intrinsic {Josephson} junctions and terahertz
  electromagnetic radiation},}\ }\href {\doibase 10.1088/0953-2048/23/5/053001}
  {\bibfield  {journal} {\bibinfo  {journal} {Superconductor Science and
  Technology}\ }\textbf {\bibinfo {volume} {23}},\ \bibinfo {pages} {053001}
  (\bibinfo {year} {2010})}\BibitemShut {NoStop}%
\bibitem [{\citenamefont {Bulaevskii}\ \emph {et~al.}(2008)\citenamefont
  {Bulaevskii}, \citenamefont {Koshelev},\ and\ \citenamefont
  {Tachiki}}]{PhysRevB.78.224519}%
  \BibitemOpen
  \bibfield  {author} {\bibinfo {author} {\bibfnamefont {L.~N.}\ \bibnamefont
  {Bulaevskii}}, \bibinfo {author} {\bibfnamefont {A.~E.}\ \bibnamefont
  {Koshelev}}, \ and\ \bibinfo {author} {\bibfnamefont {M.}~\bibnamefont
  {Tachiki}},\ }\bibfield  {title} {\enquote {\bibinfo {title} {Shapiro steps
  and stimulated radiation of electromagnetic waves due to josephson
  oscillations in layered superconductors},}\ }\href {\doibase
  10.1103/PhysRevB.78.224519} {\bibfield  {journal} {\bibinfo  {journal} {Phys.
  Rev. B}\ }\textbf {\bibinfo {volume} {78}},\ \bibinfo {pages} {224519}
  (\bibinfo {year} {2008})}\BibitemShut {NoStop}%
\bibitem [{\citenamefont {Lin}\ and\ \citenamefont
  {Hu}(2010)}]{PhysRevB.82.020504}%
  \BibitemOpen
  \bibfield  {author} {\bibinfo {author} {\bibfnamefont {Shi-Zeng}\
  \bibnamefont {Lin}}\ and\ \bibinfo {author} {\bibfnamefont {Xiao}\
  \bibnamefont {Hu}},\ }\bibfield  {title} {\enquote {\bibinfo {title}
  {Response and amplification of terahertz electromagnetic waves in intrinsic
  josephson junctions of layered high-${T}_{c}$ superconductor},}\ }\href
  {\doibase 10.1103/PhysRevB.82.020504} {\bibfield  {journal} {\bibinfo
  {journal} {Phys. Rev. B}\ }\textbf {\bibinfo {volume} {82}},\ \bibinfo
  {pages} {020504} (\bibinfo {year} {2010})}\BibitemShut {NoStop}%
\bibitem [{\citenamefont {Dienst}\ \emph {et~al.}(2013)\citenamefont {Dienst},
  \citenamefont {Casandruc}, \citenamefont {Fausti}, \citenamefont {Zhang},
  \citenamefont {Eckstein}, \citenamefont {Hoffmann}, \citenamefont {Khanna},
  \citenamefont {Dean}, \citenamefont {Gensch}, \citenamefont {Winnerl},
  \citenamefont {Seidel}, \citenamefont {Pyon}, \citenamefont {Takayama},
  \citenamefont {Takagi},\ and\ \citenamefont
  {Cavalleri}}]{dienst_optical_2013}%
  \BibitemOpen
  \bibfield  {author} {\bibinfo {author} {\bibfnamefont {A.}~\bibnamefont
  {Dienst}}, \bibinfo {author} {\bibfnamefont {E.}~\bibnamefont {Casandruc}},
  \bibinfo {author} {\bibfnamefont {D.}~\bibnamefont {Fausti}}, \bibinfo
  {author} {\bibfnamefont {L.}~\bibnamefont {Zhang}}, \bibinfo {author}
  {\bibfnamefont {M.}~\bibnamefont {Eckstein}}, \bibinfo {author}
  {\bibfnamefont {M.}~\bibnamefont {Hoffmann}}, \bibinfo {author}
  {\bibfnamefont {V.}~\bibnamefont {Khanna}}, \bibinfo {author} {\bibfnamefont
  {N.}~\bibnamefont {Dean}}, \bibinfo {author} {\bibfnamefont {M.}~\bibnamefont
  {Gensch}}, \bibinfo {author} {\bibfnamefont {S.}~\bibnamefont {Winnerl}},
  \bibinfo {author} {\bibfnamefont {W.}~\bibnamefont {Seidel}}, \bibinfo
  {author} {\bibfnamefont {S.}~\bibnamefont {Pyon}}, \bibinfo {author}
  {\bibfnamefont {T.}~\bibnamefont {Takayama}}, \bibinfo {author}
  {\bibfnamefont {H.}~\bibnamefont {Takagi}}, \ and\ \bibinfo {author}
  {\bibfnamefont {A.}~\bibnamefont {Cavalleri}},\ }\bibfield  {title} {\enquote
  {\bibinfo {title} {Optical excitation of {Josephson} plasma solitons in a
  cuprate superconductor},}\ }\href {\doibase 10.1038/nmat3580} {\bibfield
  {journal} {\bibinfo  {journal} {Nature Materials}\ }\textbf {\bibinfo
  {volume} {12}},\ \bibinfo {pages} {535--541} (\bibinfo {year}
  {2013})}\BibitemShut {NoStop}%
\bibitem [{\citenamefont {Rajasekaran}\ \emph {et~al.}(2016)\citenamefont
  {Rajasekaran}, \citenamefont {Casandruc}, \citenamefont {Laplace},
  \citenamefont {Nicoletti}, \citenamefont {Gu}, \citenamefont {Clark},
  \citenamefont {Jaksch},\ and\ \citenamefont
  {Cavalleri}}]{rajasekaran_parametric_2016}%
  \BibitemOpen
  \bibfield  {author} {\bibinfo {author} {\bibfnamefont {S.}~\bibnamefont
  {Rajasekaran}}, \bibinfo {author} {\bibfnamefont {E.}~\bibnamefont
  {Casandruc}}, \bibinfo {author} {\bibfnamefont {Y.}~\bibnamefont {Laplace}},
  \bibinfo {author} {\bibfnamefont {D.}~\bibnamefont {Nicoletti}}, \bibinfo
  {author} {\bibfnamefont {G.~D.}\ \bibnamefont {Gu}}, \bibinfo {author}
  {\bibfnamefont {S.~R.}\ \bibnamefont {Clark}}, \bibinfo {author}
  {\bibfnamefont {D.}~\bibnamefont {Jaksch}}, \ and\ \bibinfo {author}
  {\bibfnamefont {A.}~\bibnamefont {Cavalleri}},\ }\bibfield  {title} {\enquote
  {\bibinfo {title} {Parametric amplification of a superconducting plasma
  wave},}\ }\href {\doibase 10.1038/nphys3819} {\bibfield  {journal} {\bibinfo
  {journal} {Nature Physics}\ }\textbf {\bibinfo {volume} {12}},\ \bibinfo
  {pages} {1012--1016} (\bibinfo {year} {2016})}\BibitemShut {NoStop}%
\end{thebibliography}%

\end{document}